\documentstyle [12pt,epsf] {article}

\catcode`@=12
\def\be{\begin{equation}}
\def\ee{\end{equation}}
\def\lsim{\mathrel{\vcenter{\hbox{$<$}\nointerlineskip\hbox{$\sim$}}}}
\newcommand{\bea}{\begin{eqnarray}}
\newcommand{\eea}{\end{eqnarray}}
\newcommand{\nn}{\nonumber}

\begin{document}
\vspace{0.5in}
\oddsidemargin -.375in
\newcount\sectionnumber
\sectionnumber=0

\def\lsim{\mathrel{\vcenter{\hbox{$<$}\nointerlineskip\hbox{$\sim$}}}}
\thispagestyle{empty}


\vskip0.5truecm

\begin{center}

{\large \bf
\centerline{Probing new physics in  $ B \to \phi \pi $ decay }}

\vspace*{1.0cm}
{ Anjan K. Giri$^1$ and  Rukmani Mohanta$^{1,2}$ } \vskip0.3cm
{\it  $^1$ Physics Department, Technion-Israel 
Institute of Technology, 32000 Haifa, Israel}\\
{\it $^2$ School of Physics, University of Hyderabad,
Hyderabad - 500046, India
} \\
\vskip0.5cm
\bigskip
(\today)
\vskip0.5cm

\begin{abstract}
We analyze the rare decay mode $B \to \phi \pi $, which is a
pure penguin induced process, receiving dominant contribution
from the electroweak penguins. Thus the standard model branching ratio 
is expected to be very small, which makes it as a  
sensitive probe of new physics. Using QCD factorization approach,
we find the branching ratio in the standard model as
${\rm Br}(B^- \to \phi \pi^-)\simeq 5 \times 10^{-9}$. 
Exploring some of the beyond standard model scenarios the branching
ratio is found to 
be $\sim {\cal O}(10^{-8})$ in the minimal supersymmetric standard
model (MSSM) with mass insertion approximation and in the extended
technicolor model (TC2). The existence of an    
 extra vector-like down quark (VLDQ) model predicts it to be
 $\sim {\cal O}(10^{-7})$. The recent BaBar result
on $B^0 \to \phi \pi^0$ might be a strong
indication of new physics effect present in the penguin induced process
$B \to \phi \pi$.\\

\vspace*{0.25 truecm}
 
\end{abstract}
\end{center}

\thispagestyle{empty}
\newpage
\baselineskip=14pt

\section{Introduction}

The intensive search for physics beyond the standard model (SM) is
performed now a days in various areas of particle physics. In this respect, 
the $B$ meson system can also be used as a complementary probe to the
search for new physics. The main objectives of the ongoing and the 
future $B$ factory experiments are to explore in detail the origin of 
CP violation,  to test the standard model
at an unexpected level of precision and to look for possible existence of
new physics effects.  In the $B$ experiments new physics beyond the SM
may manifest itself in various ways, e.g., {\it (i)}
decays which are expected to be rare in the SM and are found to have
large branching ratios {\it (ii)} CP violating rate asymmetries
which are expected to vanish or to  be very small in the SM are
found to be significantly large {\it (iii)} the discrepancy between
the mixing induced CP asymmetry between various $B$ decay processes which are
dominated by a single decay amplitude with same weak phase,
i.e., $S_{\psi K_S}$ and $S_{\phi K_S}$ etc.

Thus the rare $B$ meson decays are suggested to give good opportunities 
for discovering new physics beyond the SM. The discrepancy between
the recently measured $S_{\phi K_S}$ and $S_{\psi K_S}$ 
\cite{belle0, tom1} has already
given an indication of the possible existence of new physics in the
$B$ decay amplitudes (i.e., in the penguin induced $B\to \phi K_S$ decay).

In this paper, we would like to explore the presence of
new physics in another pure penguin induced decay mode $B\to \phi \pi$. 
It proceeds through the quark level transition $b \to d \bar s s$, which is
a flavor changing neutral current (FCNC) process at the one loop level.
The interesting
feature of this process is that it is dominated by electroweak
penguin contribution. The QCD penguins should play a minor role
for this transition since $\bar s$ and $s$ quarks emerging from
the gluons of the usual QCD penguin diagram form a color
octet state and consequently cannot build up the
$\phi $ meson which is a $\bar s s$ color singlet state. Therefore the SM
prediction for the branching ratio for this process is expected to be
quite small. 
Thus as this mode is highly suppressed in the SM, it may serve 
as a good hunting ground to look for new
physics beyond the SM.

Recently, these decay modes have been searched for by 
the BaBar collaboration \cite{babar03, babar04}:   
\bea
{\rm Br}(B^\pm \to \phi \pi^\pm) & < & 0.41 \times 10^{-6}\;,\nn\\
{\rm Br}(B^0 \to \phi \pi^0) &=& (0.2_{-0.3}^{+0.4} \pm 0.1) \times 
10^{-6}\;,\nn\\
& < & (1.2 \pm 0.8) \times 10^{-6}\;.\label{eq:expt}
\eea
On the theoretical side the decay mode $B^- \to \phi \pi^- $
has been studied recently
in Ref. \cite{cheng03}.  Using QCD factorization approach 
the branching ratio has been obtained in the SM and in the constrained 
minimal supersymmetric 
standard model. It has also been studied in Ref.
\cite{eilam03}  in the SM and R-parity violating sypersymmetric model and 
in the Refs. \cite{du02, fleis94} using the standard model approach.

In this paper,  we will first reanalyze this
process again in the SM for the sake of completeness,
using QCD factorization. It should also be noted that there is 
not good agreement of the SM predictions in the literature.
We then consider the 
minimal supersymmetric model with mass insertion approximation. 
Basically we will
confine ourselves to the case where the new contributions to the
$B \to \phi \pi$ will arise from the gluino mediated $b \to d \bar s s$
process induced by the flavor mixing in the down s-quark sector.
Next we will calculate the branching ratio in the framework of topcolor 
assisted technicolor model (TC2)  and in the model with
an extra vector like down quark (VLDQ).

\section{Standard Model contribution}

In the SM, the decay process $B \to \phi \pi$ receives 
contribution from the
quark level transition $b \to d \bar s s$, which is induced by the
pure penguin diagram with dominant contributions coming from
electroweak penguins. The effective Hamiltonian describing the
decay $b\to d\bar ss$ \cite{beneke99} is given as
\be 
{\cal H}_{eff}^{SM}=
\frac{G_F}{\sqrt{2}}\biggr[
V_{qb}V_{qd}^* \sum_{i=3}^{10}C_i O_i
 \biggr],
\ee
where $q=u,~c$. $O_3, \cdots, O_{6}$ and $O_7,
\cdots, O_{10}$ are the standard model QCD and electroweak penguin operators,
respectively.
The values of the Wilson coefficients at the scale $\mu \approx m_b$
in the NDR scheme are given
in Ref. \cite{buca96} as
\bea
&&C_3=0.014\;,~~~~~~~C_4=-0.035\;,~~~~~~C_5=0.009\;,~~~~~C_6=-0.041\;,
\nn\\
&&C_7=-0.002\alpha\;,~~~C_8=0.054 \alpha\;,
 ~~~C_9=-1.292\alpha\;,~~~~C_{10}=0.263 \alpha\;.
\eea
We use QCD factorization \cite{beneke99}  
to evaluate the hadronic matrix elements.
In this method, the decay amplitude can be represented in the form
\be
\langle \phi \pi^- |O_i |B^- \rangle =
\langle \phi \pi^- |O_i |B^- \rangle_{\rm fact}\biggr[
1+\sum r_n \alpha_s^n + {\cal O}(\Lambda_{\rm QCD}/m_b)
\biggr]\;,
\ee
where $\langle \phi \pi^- |O_i |B^- \rangle_{\rm fact}$ denotes the 
naive factorization result and $\Lambda_{\rm QCD} \sim 225$ MeV, 
the strong interaction scale. The second and third terms in the square
bracket represent higher order $\alpha_s$ and $\Lambda_{\rm QCD}/m_b$
corrections to hadronic matrix elements.

In the heavy quark limit the decay amplitude for the $B^- \to \phi \pi^- $
process, arising from the penguin diagrams, is given  as 
\bea
A^{SM}(B^- \to \phi \pi^-) = \frac{G_F}{\sqrt 2}\sum_{q=u,c} 
V_{qb}V_{qd}^* \biggr[a_3^q+a_5^q-\frac{1}{2}\left (
a_7^q+a_9^q \right ) \biggr]X\;,\label{eq:sm}
\eea
where $X$ is the factorized matrix element. The amplitude for
$B^0 \to \phi \pi^0$ is related to $B^- \to \phi \pi^-$ by
\bea
A(B^0 \to \phi \pi^0)= \frac{1}{\sqrt 2}
A(B^- \to \phi \pi^-)\;.
\eea
Using the form factors and decay
constants, defined as \cite{bsw}
\bea
\langle \pi^-(p_\pi) |\bar d \gamma ^\mu b | B^-(p_B) \rangle &=&
\biggr[(p_B+p_\pi)^\mu-\frac{m_B^2-m_\pi^2}{q^2}\biggr] F_1(q^2)\nn\\
&+&
\frac{m_B^2-m_\pi^2}{q^2}q^\mu F_0(q^2)\;,\nn\\
\langle \phi(q, \epsilon) |\bar s \gamma ^\mu s | 0 \rangle &=&
f_\phi~ m_\phi~ \epsilon^\mu\;,
\eea
we obtain the factorized matrix element $X$ as 
\bea
 X &=& \langle  \pi^- (p_\pi)| \bar d
\gamma_\mu(1-\gamma_5)b | B^-(p_B) \rangle
\langle \phi(q, \epsilon )|\bar s
\gamma^\mu(1-\gamma_5)s|0 \rangle \nn\\
& = & 2 F^{B \to \pi}_1(m_\phi^2)~f_\phi~ m_\phi~
(\epsilon \cdot p_B)\;.
\eea
The coefficients $a_i^q$,s which contain next to leading order
(NLO) and hard scattering corrections are given as \cite{du02, yang00}
\bea
a_3^u &=& a_3^c ~=~ C_3+\frac{C_4}{N}+\frac{\alpha_s}{4 \pi}\frac{C_F}{N}
C_4 F_\phi \;,\nn\\
a_5^u &=& a_5^c ~=~ C_5+\frac{C_6}{N}+\frac{\alpha_s}{4 \pi}
\frac{C_F}{N}C_6(- F_\phi -12) \;,\nn\\
a_7^u &=& a_7^c ~=~ C_7+\frac{C_8}{N}+\frac{\alpha_s}{4 \pi}
\frac{C_F}{N}C_8(- F_\phi -12) \;,\nn\\
a_9^u &=& a_9^c ~=~ C_9+\frac{C_{10}}{N}+\frac{\alpha_s}{4 \pi}
\frac{C_F}{N}C_{10} F_\phi 
\;,\label{qcd}
\eea
where  $N=3$, is the number of
colors and $C_F=(N^2-1)/2N$. 
The  parameters present in (\ref{qcd}) are given as
\bea
F_\phi &=& -12 \ln \frac{\mu}{m_b}-18+f_\phi^I +f_\phi^{II}\;,\nn\\
f_\phi^I &= & \int_0^1 dx~ g(x) \phi_\phi(x)\;,\nn\\
g(x) &= & 3 \frac{1-2x}{1-x}\ln x -3i \pi\;, \nn\\
f_\phi^{II} &= & \frac{4 \pi^2}{N} \frac{f_\pi f_B}{F_1^{B \to \pi}(0) m_B^2}
\int_0^1 \frac{dz}{z} \phi_B(z) \int_0^1 \frac{dx}{x} \phi_\pi(x) 
\int_0^1 \frac{dy}{y} \phi_\phi(y)\;.
\eea
The light cone distribution amplitudes (LCDA's) at twist two order 
are given as
\bea
&&\phi_B(x) = N_B x^2(1-x)^2 {\rm exp}\biggr(-\frac{m_B^2 x^2}{2
\omega_B^2}\biggr) \;,\nn\\
&&\phi_{\pi, \phi}(x)=  6x(1-x)\;,
\eea
where $N_B$ is the normalization factor satisfying
$ \int_0^1 dx \phi_B(x)=1$ and $\omega_B=0.4$ GeV. 
The branching ratio can be obtained using the formula
\bea
{\rm Br}(B^- \to \phi \pi^-) & = & \tau_{B^-} \frac{|p_{\rm cm}|^3}{
8 \pi m_\phi^2}~|{A(B^- \to \phi \pi^-)}/({\epsilon \cdot p_B})|^2\;,\nn\\
{\rm Br}(B^0 \to \phi \pi^0) &=& \frac{\kappa}{2} ~
{\rm Br}(B^- \to \phi \pi^-)\;, 
\eea
where $\kappa= \tau_{B^0}/\tau_{B^-}$ and 
$p_{\rm cm}$ is the momentum of the outgoing particles in the $B$
meson rest frame. 

For numerical evaluation we have used the following input parameters. 
The value of the form factor 
at zero recoil is taken as $F_1^{B \to \pi}(0)=$ 0.33, 
and its value at $q^2=m_\phi^2$ can be obtained 
using simple pole dominance ansatz \cite{bsw} as 
$F_1^{B \to \pi}(m_\phi^2)=$ 0.34. 
The  values of the decay constants are as $f_{\phi}=$ 0.233 GeV, 
$f_B=0.19 $ GeV,
$f_\pi$=0.131 GeV, the particle masses and the lifetime of 
$B$ mesons $\tau_{B^-}=
1.674 $ ps, $\tau_{B^0}=1.542$ ps are taken from \cite{pdg}. 
For the CKM matrix elements,
we have used the 
Wolfenstein parameterization and have
taken the values of the parameters
$A=0.819 \pm 0.040$, $\lambda=0.2237 \pm 0.0033$, 
$\rho=0.224\pm 0.039$ and $\eta=0.324 \pm 0.039$.
With these input parameters,  we obtain
the branching ratio for $B^- \to \phi \pi^-$ in the SM as
\bea
{\rm Br}(B^- \to \phi \pi^-)|_{SM} &= &(5.5 \pm 0.9) 
\times 10^{-9}\;,\nn\\
{\rm Br}(B^0 \to \phi \pi^0)|_{SM} &=&(2.5 \pm 0.4) 
\times 10^{-9}\;.
\eea
These predicted values are quite below the present experimental upper
limit and the central value of $B^0 \to \phi \pi^0$ (\ref{eq:expt}).
It should be noted that our prediction is in agreement with \cite{cheng03},
the slight difference is due to the difference in the used CKM 
parameters and formfactor.


\section{Supersymmetric contribution}


Now we study the decay process $B \to \phi \pi $, in the minimal
supersymmetric standard (MSSM) model with gluino contributions, because
the chargino and charged Higgs contributions are expected to be suppressed
by the small electroweak gauge couplings. Thus, the one loop contributions
to the above mentioned decay process can be induced by s-quark and
gluino penguin and box diagrams. These gluino mediated FCNC contributions are 
of the order of strong interaction strength, which may exceed the
existing limits.  It is customary to rotate the effects, so that they occur
in s-quark propagator rather than in couplings and to parametrize them
in terms of dimensionless parameters. Here we work in the usual mass insertion
approximation \cite{hall86, gabb96} where the flavor mixing
$i \to j$ in the down-type squarks associated with $\tilde q_B$ and
$\tilde q_A$ are parametrized by  $(\delta^d_{ij})_{AB}$, with
$A,~B=L,~R$ and $i,j$ as the generation indices. More explicitly
$(\delta^d_{LL})_{ij}
=({V_L^d}^\dagger M_{\tilde d}^2 V_L^d)_{ij}/ m_{\tilde q}^2$, where 
$M_{\tilde d}^2$
is the squared down squark mass matrix and $m_{\tilde q}$ is
the average squark mass.
$V_d$ is the matrix which diagonalizes the down quark mass matrix.

The new effective $\Delta B=1$ Hamiltonian relevant for the $B \to \phi \pi$
process arising from new penguin/box diagrams with gluino-squark in the 
loops is given as
\be
{\cal H}_{eff}^{SUSY} = -\frac{G_F}{\sqrt 2} V_{tb}V_{td}^*
\sum_{i=3}^6 \left [ C_i^{NP}O_i+ \tilde C_i^{NP} \tilde O_i \right ]\;,
\ee
where $O_i$ are the QCD penguin operators and the $C_i^{NP}$ are the 
new Wilson coefficients. The operators $\tilde O_i$ 
are obtained from $O_i$  by exchanging $L \leftrightarrow
R$. Thus including the new physics contribution, one can write
the total amplitude for $B \to \phi \pi$ process as
\bea
A^{T}=A^{SM}\left [1+\left |\frac{A^{SUSY}}{A^{SM}}\right |
e^{i \theta} \right ]\;,
\eea
where $A^{SUSY}$ is the new physics amplitude arising from 
minimal supersymmetric model and $\theta$ is the relative phase
 between the
SM and the new physics decay amplitudes. The corresponding branching
ratio is given as
\be
{\rm Br}={\rm Br}^{SM}\left [1+2 \cos \theta |{A^{SUSY}}/{A^{SM}}|+
 |{A^{SUSY}}/{A^{SM}}|^2 \right]\;,\label{eq:br}
\ee
where ${\rm Br}^{SM}$ is the SM branching ratio. 
To evaluate the
amplitude in the MSSM, we have to first evaluate the Wilson coefficients 
at the $b$ quark mass scale.
At the leading order in mass insertion approximation the new Wilson 
coefficients corresponding to each of the operator at the scale
$\mu \sim \tilde m \sim M_W$ are given as \cite{gabb96}
\bea 
C_3^{NP} & \simeq & -\frac{\sqrt 2 \alpha_s^2}{4 G_F V_{tb}
V_{td}^* m_{\tilde q}^2}\left ( \delta_{LL}^d \right )_{13}
\left [ - \frac{1}{9}B_1(x) -\frac{5}{9} B_2(x)-\frac{1}{18}P_1(x)
-\frac{1}{2}P_2(x) \right ]\;,\nn\\
C_4^{NP} & \simeq & -\frac{\sqrt 2 \alpha_s^2}{4 G_F V_{tb}
V_{td}^* m_{\tilde q}^2}\left ( \delta_{LL}^d \right )_{13}
\left [ - \frac{7}{3}B_1(x) +\frac{1}{3} B_2(x)-\frac{1}{6}P_1(x)
+\frac{3}{2}P_2(x) \right ]\;,\nn\\
C_5^{NP} & \simeq & -\frac{\sqrt 2 \alpha_s^2}{4 G_F V_{tb}
V_{td}^* m_{\tilde q}^2}\left ( \delta_{LL}^d \right )_{13}
\left [  \frac{10}{9}B_1(x) +\frac{1}{18} B_2(x)-\frac{1}{18}P_1(x)
-\frac{1}{2}P_2(x) \right ]\;,\nn\\
C_6^{NP} & \simeq & -\frac{\sqrt 2 \alpha_s^2}{4 G_F V_{tb}
V_{td}^* m_{\tilde q}^2}\left ( \delta_{LL}^d \right )_{13}
\left [ - \frac{2}{3}B_1(x) +\frac{7}{6} B_2(x)+\frac{1}{6}P_1(x)
+\frac{3}{2}P_2(x) \right ]\;.
\eea
The corresponding $ \tilde C_i$ are obtained from $C_i^{NP}$ by
interchanging $L \leftrightarrow R$. The functions appear in these
expressions can be found in Ref. \cite{gabb96} 
and $x=m_{\tilde g}^2/m_{\tilde q}^2 $. Because the parity of the
vector meson $\phi$ is opposite to that of $B$ and $\pi $ mesons,
which are pseudoscalars, the gluino loop effects appear as
$\left ( \delta_{LL}^d \right )_{13}+\left ( \delta_{RR}^d \right )_{13}$.
The Wilson coefficients at low  energy
$C_i^{NP}(\mu)$, $\mu\sim {\cal O}(m_b)$ can be obtained from $C_i^{NP}(M_W)$
by using the Renormalization Group (RG) equation as discussed in
Ref. \cite{buca96}, as
\be
{\bf C}(\mu) ={\bf U}_5(\mu, M_W) {\bf C}(M_W)\;,
\ee
where ${\bf C}$ is the $6 \times 1$ column vector of the
Wilson coefficients and
${\bf U}_5(\mu, M_W)$ is the five-flavor $6 \times 6$ evolution matrix.
In the next-to-leading order (NLO), ${\bf U}_5(\mu, M_W)$ is given by
\be
{\bf U}_5(\mu, M_W)=\left (1+\frac{\alpha_s(\mu)}{4 \pi} {\bf J} \right )
{\bf U}_5^{(0)}(\mu, M_W)\left (1-\frac{\alpha_s(M_W)}{4 \pi} {\bf J} 
\right )\;,
\ee
where ${\bf U}_5^{(0)}(\mu, M_W)$ is the leading order (LO)
evolution matrix and ${\bf J}$ denotes the NLO corrections to the evolution. 
The explicit forms of ${\bf U}_5(\mu, M_W)$ and 
${\bf J}$ are given in Ref. \cite{buca96}. 

For the numerical analysis, we fix the SUSY parameter as $m_{\tilde q}
=m_{\tilde g}=500 $ GeV, $\alpha_s(M_W)=0.118$,
$\alpha_s(m_b=4.4~ GeV)=0.221$. Thus, the values of the Wilson coefficients 
evaluated at the $b$ quark mass scale are given as (where the
common factor $\left ( \delta_{LL}^d \right )_{13}$ has been factored
out)
\bea
C_3^{NP}=0.025\;,~~~~
C_4^{NP}=-0.021\;,~~~~
C_5^{NP}=-0.003\;,~~~~
~C_6^{NP}=-0.087\;.
\eea
Now evaluating the matrix elements of the operators $O_{3-6}$
as done in Eqs. (\ref{eq:sm})-(\ref{qcd}) 
for the SM i.e., replacing the SM Wilson
coefficients $C_{3-6}$ by their corresponding SUSY counterparts,
we obtain the total amplitude in the MSSM as
\be
A(B^- \to \phi \pi^-)= A^{SM}\biggr(1+ 2.525  \left ((\delta^d_{13})_{LL}
+(\delta^d_{13})_{RR} \right )\biggr)\;.
\ee
The constraint  on $ (\delta^d_{13})_{LL}$ is obtained from
$B^0 -\bar B^0$ mixing, however for 
  $(\delta^d_{13})_{RR}$, it is  not very stringent. We use 
the conservative limit for $m_{\tilde q}$=500 GeV and $x=1$ as
$(\delta^d_{13})_{LL} < \sqrt{|{\rm Re}(\delta^d_{13})_{LL}^2|}
\sim 9.8 \times 10^{-2}$ \cite{gabb96},
$ (\delta^d_{13})_{LL} \sim  (\delta^d_{13})_{RR}$ and their weak phases 
to be equal.
Thus, we obtain the branching
ratio in MSSM as
\bea
{\rm Br}(B^-\to \phi \pi^- )|_{MSSM} &\leq & 1.2 \times 10^{-8}\;,\nn\\
{\rm Br}(B^0\to \phi \pi^0)|_{MSSM}& \leq & 0.5 \times 10^{-8}\;.
\eea
Although the predicted branching ratios in MSSM are enhanced by one
order from their SM values but they are still well below the present
upper limits.


\section{Contribution from the VLDQ Model}


Now we consider the model with an additional vector like down
quark \cite{gross98}. It is a model with an extended quark sector.
In addition to the three standard generation of quarks, there is an
$SU(2)_L$  singlet of charge $-1/3$. The mixing of these singlet quarks
with the three SM down type quarks provides a framework 
to study the deviation from unitarity constraint of $3 \times 3$ CKM matrix. 
The important feature of this model is that it allows $Z$-mediated FCNC
at the tree level. Thus,
the presence of an additional singlet down quark implies 
a $4 \times 4$ matrix $V_{i \alpha}$
$(i=u,c,t,4,~\alpha= d,s,b,b')$, diagonalizing the down quark mass
matrix. For our purpose, the relevant information for the low
energy physics is encoded in the extended mixing matrix. The
charged currents are unchanged except that the $V_{CKM}$ is now the $3
\times 4$  upper submatrix of $V$. However, the distinctive feature
of this model is that FCNC enters the neutral current Lagrangian of
the left handed downquarks :
 \be
 {\cal L}_Z= \frac{g}{2 \cos
\theta_W} \left [ \bar u_{Li} \gamma^{\mu} u_{Li} - \bar d_{L
\alpha}U_{\alpha \beta} \gamma^\mu d_{L \beta}-2 \sin^2 \theta_W
J_{em}^\mu \right ]Z_{\mu}\;,
\ee
with
\be
U_{\alpha \beta}
=\sum_{i=u,c,t} V_{\alpha i}^\dagger V_{i \beta} =\delta_{\alpha
\beta} - V_{4 \alpha}^* V_{4 \beta}\;,
\ee
where $U$ is the neutral
current mixing matrix for the down sector which is given above. As
$V$ is not unitary, $U \neq {\bf{1}}$. In particular its
non-diagonal elements do not vanish :
\be
U_{\alpha \beta}= -V_{4
\alpha}^* V_{4 \beta} \neq 0~~~{\rm for}~ \alpha \neq \beta\;.
\ee
Since the various $U_{\alpha \beta}$ are non vanishing they would
signal new physics and the presence of FCNC at the tree level, this
can substantially enhance the branching ratio of $B \to \phi \pi $. 
The observed discrepancy between $S_{\phi K_S}$ and $S_{\psi K_S}$
can be explained in this model \cite{giri03}. 
The
new element $U_{d b}$ which is relevant to our study is given as
\be
U_{db}= V_{ud}^* V_{ub}+V_{cd}^*V_{cb}+V_{td}^*V_{tb}\;.
\ee
Thus the decay modes $B \to \phi \pi $ receive the new 
contributions from the  $Z$-mediated FCNC transitions and
the new additional operator is given as
\bea
O^{VLDQ}=[\bar d_\alpha \gamma^\mu (1-\gamma_5) b_\alpha][\bar s_\beta
\gamma_\mu(C_V^s -C_A^s \gamma_5) s_\beta] \;,\nn\\
\eea
where $C_V^s$ and $C_A^s$ are the vector and axial vector $Z s \bar s$
couplings. Using the identity
$(C_V^s-C_A^s \gamma_5)=[(C_V^s+C_A^s)(1- \gamma_5)+(C_V^s-C_A^s)(1+ \gamma_5)]
/2$, the effective Hamiltonian for $B \to \phi \pi$ transition
is given as 
\bea
{\cal H}^{VLDQ}_{eff} &=& \frac{G_F}{2 \sqrt 2}U_{db}\left [ (C_V^s+C_A^s) 
(\bar d b)_{V-A}(\bar s s)_{V-A}
+(C_V^s- C_A^s) (\bar d b)_{V-A}(\bar s s)_{V+A}\right ]\;,
\eea
where the subscripts in the currents denote the usual $(V-A)$ and $
(V+A)$ currents. Now evaluating the matrix elements of the operators
the transition amplitude for the process in VLDQ model is given as
\be
A^{VLDQ}(B^- \to \phi \pi^-) = \frac{G_F}{2\sqrt 2} U_{db}
X
(\epsilon \cdot p_B)\left ( (C_V^s+C_A^s) (1+\frac{\alpha_s}{4\pi}F_\phi) 
+
(C_V^s-C_A^s) \right )\;,
\ee
where we have also included the leading order nonfacorizable contributions.
Now using the value for $C_V^s$ and $C_A^s$  as
\be
C_V^s= -\frac{1}{2}+\frac{2}{3} \sin^2 \theta_W\;,~~~~~C_A^s=-\frac{1}{2}\;,
\ee
 $\sin^2 \theta_W$=0.23, alongwith
$|U_{db}| \leq 1.2  \times 10^{-3}$ \cite{boim01}, we obtain the ratio 
of VLDQ and SM amplitudes as
\be
|\frac{A^{VLDQ}(B^- \to \phi \pi^-)}
{A^{SM}(B^- \to \phi \pi^-)}|\leq 8.16
\;.
\ee
The branching ratio in VLDQ model obtained using Eq. (\ref{eq:br}) as
\bea
{\rm Br}( B^- \to \phi \pi^-)|_{VLDQ} &\leq & 4.6  \times 10^{-7}\nn\\
{\rm Br}( B^0 \to \phi \pi^0)|_{VLDQ} &\leq & 2.1  \times 10^{-7}\;,
\eea
which are two order above the standard model prediction and the branching 
ratio for $B^0 \to \phi \pi^0$ is in agreement with the present data
(\ref{eq:expt}).


\section{Contribution from the TC2 Model}

Now we calculate the branching ratio for $B \to \phi \pi$ process
in the framework of the topcolor assisted technicolor model (TC2) 
\cite{hill95, lane95}.
It is well known that technicolor is one of the important 
candidates for the
electroweak symmetry breaking and the extended technicolor 
model was proposed to generate the ordinary fermion masses. 
In order to generate large top quark mass the topcolor model has been 
constructed  recently. Apart from some difference in group 
structure and/or particle contents, all TC2 models have similar
common features.  Following the TC2 model of Hill \cite{hill95},
the charmless decays $B \to PP,~PV$ are studied in Ref \cite{xiao01},
using the generalized factorization approach.
In this paper we will also use the same procedure to evaluate
the new physics contribution to the $B \to \phi \pi$ mode, 
but we will use QCD factorization approach
to evaluate the hadronic matrix elements.

In TC2 model, there exist top-pions ($ \tilde \pi^\pm $ and
$\tilde \pi^0 $), charged and neutral $b$-pions
($\tilde H^\pm $, $\tilde H^0$ and $\tilde A^0 $ ) and 
technipions ($\pi_1^\pm$ and $\pi_8^\pm$).  The coupling of
top-pions to $t$- and $b$-quarks can be written as
\be
\frac{m_t^*}{F_{\tilde \pi}}\biggr[
i \bar t t \tilde \pi^0 + + i \bar t_R b_L \tilde \pi^+
+\frac{m_b^*}{m_t^*}\bar t_L b_R \tilde \pi^+ + 
{\rm h.c.}\biggr]\;,
\ee
where $m_t^*=(1-\epsilon)m_t$ and $m_b^* \sim $1 GeV denote
the masses of top and bottom quarks generated by the topcolor 
interactions and $F_{\tilde \pi}$ is the top-pion decay
constant. At low energy, potentially large FCNC
arise when the quark fields are rotated from their weak 
eigenbasis to their mass eigenbasis, realized by the matrices
$U_{L,R}$ for the up-type quarks  and $D_{L,R}$ for the down type quarks.
Thus for example, making the replacement 
\bea
b_L  &\to & D_L^{bd} d_L +D_L^{bs} s_L + D_L^{bb}b_L\;,\nn\\
b_R &\to & D_R^{bd} d_R + D_R^{bs}s_R + D_R^{bb}b_R\;,
\eea
the FCNC interactions will be induced. In the TC2 model the corresponding 
flavor changing effective Yukawa couplings are 
\be
\frac{m_t^*}{F_{\tilde \pi}}\biggr[ i \tilde \pi^+
\left (D_L^{bs} \bar t_R s_L +D_L^{bd} \bar t_R d_L
\right ) + i \tilde H^+\left (D_R^{bs} \tilde t_L s_R
+D_R^{bd} \bar t_L d_R \right )+ {\rm h.c.} \biggr]\;.
\ee
The constraints on the parameters in the TC2 model are discussed in detail 
in Ref \cite{xiao01}, obtained from various experimental data. For the mixing
matrices the `` square root ansatz'' is considered, i.e.
 $D_L^{bd}=V_{td}/2$ and $D_L^{bs}=V_{ts}/2$. The other parameters are 
given as
\bea
m_{\pi_1}=100~GeV, ~~m_{\pi_8}=200~GeV,~~~F_{\tilde \pi}=50~ GeV,
~~F_\pi=120 ~GeV~~\epsilon=0.05\;,
\eea
where $F_\pi$ is the technipion decay constant.

In this model, the decay process $b \to d \bar s s$ is induced by the
exchange of the charged top pions $ \tilde \pi^{\pm}$ and technipions 
$ \pi_1^\pm$ and $\pi_8^\pm$ through the strong and electroweak
penguin diagrams.  Combining the new physics contributions with
their SM counterparts, the effective Wilson coefficients are evaluated.

Thus the new strong and electroweak penguin diagrams can be obtained
from the corresponding penguin diagrams in the SM by replacing the internal
$W^\pm$ lines with the charged top-pions and technipions. The analytic 
expressions for these diagrams are calculated in the dimensional 
regularization with $\overline{\rm MS}$ renormalization scheme in Ref.
\cite{xiao01}.
The new contributions to the Wilson coefficients 
arising from these diagrams are given as
\bea
C_0^{TC2} &=& \frac{1}{\sqrt 2 G_F M_W^2} \left [
\frac{m_{\tilde \pi}^2}{4 F_{\tilde \pi}^2}T_0(y_t)
+  \frac{m_{ \pi_1}^2}{3 F_{\pi}^2}T_0(z_t)
+  \frac{8m_{ \pi_8}^2}{3 F_{\pi}^2}T_0(\xi_t)\right ]\;,\nn\\
D_0^{TC2} &=& \frac{1}{\sqrt 2 G_F } \left [
\frac{1}{4 F_{\tilde \pi}^2}F_0(y_t)
+  \frac{1}{3 F_{\pi}^2}\biggr(F_0(z_t)
+  8 F_0(\xi_t)\biggr)\right ]\;,\nn\\
E_0^{TC2} &=& \frac{1}{\sqrt 2 G_F } \left [
\frac{1}{4 F_{\tilde \pi}^2}I_0(y_t)
+  \frac{1}{3 F_{\pi}^2}\biggr(I_0(z_t)
+  8 I_0(\xi_t)+9 N_0(\xi_t)\biggr)\right ]\;,\nn\\
{E_0'}^{TC2} &=& \frac{1}{2\sqrt 2 G_F } \left [
\frac{1}{4 F_{\tilde \pi}^2}K_0(y_t)
+  \frac{1}{3 F_{\pi}^2}\biggr(K_0(z_t)
+  8 K_0(\xi_t)+9 L_0(\xi_t)\biggr)\right ]\;,
\eea
where $y_t={m_t^*}^2/m_{\tilde \pi}^2$ with $m_t^*=(1- \epsilon)m_t$,
$z_t=(\epsilon m_t)^2/m_{\pi_1}^2$ and $\xi_t=(\epsilon m_t)^2/m_{\pi_8}^2$.
The loop functions $T_0(x),~ F_0(x),~I_0(x),~K_0(x),~L_0(x),~N_0(x)$
are found from Ref \cite{xiao01}. 
Using the top pion and technipon masses to be 200 GeV,
the values of the $C_0,~D_0, \cdots $ functions at the $M_W$ mass scale
are obtained as \cite{xiao01}
\bea
\{C_0,~D_0,~,E_0,~E_0'\}^{TC2}|_{\mu=M_W}=\{1.27,~0.27,~0.66,-1.58\}\;.
\eea

Now combining these values with the corresponding SM values at the $W$-boson
mass
scale (
$\{C_0,~D_0,~,E_0,~E_0'\}^{SM}|_{\mu=M_W}=\{0.81,-0.48,~0.27,~0.19\}$)
, and running the resulting contributions to the $b$ quark mass scale
$(\mu=2.5$ GeV) the effective Wilson coefficients are obtained as
\cite{xiao01}
\bea
&&C_3=0.0195\;,~~~~~~C_4=-0.0441\;,~~~~~~
C_5=0.0111\;,~~~~~~C_6=-0.0535\;,\nn\\
&&C_7=0.0026\;,~~~~~~C_8=0.0018\;,~~~~~~~
 C_9=-0.0175\;,~~~~~~C_{10}=0.0049\;.
\eea
Now substituting these values in (\ref{eq:sm}) the branching ratios in the
TC2 model are found to be
\bea
{\rm Br}(B^- \to \phi \pi^-)|_{TC2} &= &1.2 \times 10^{-8}\;,\nn\\
{\rm Br}(B^0 \to \phi \pi^0)|_{TC2} &= &0.5 \times 10^{-8}\;.
\eea
Although the branching ratios are one order higher than the
corresponding SM value but they are well below the present experimental 
limits.

\section{Conclusion}


In this paper, we have analyzed the decay mode $B \to \phi \pi$ both in the
standard model and some beyond standard model scenarios. This decay mode
proceeds through the quark level FCNC transition $b \to d \bar s s$,
receiving contributions only from one-loop penguin diagrams. However, 
because of OZI suppression, the 
QCD penguins play only a minor role in this case and the dominant
contributions coming from the electroweak penguins. Therefore, in the
standard model these decays are highly suppressed, which makes them
a very sensitive probe for new physics.

Using QCD factorization approach, we found the branching ratios in the SM for
these modes as $\sim {\cal O}(10^{-9})$, which are quite below the present
experimental upper limits  $ {\cal O}(10^{-6})$. We have also calculated
the branching ratios in the MSSM
with mass insertion approximation, in the VLDQ model and in the topcolor
assisted technicolor model. The branching ratios obtained in the
MSSM and TC2 models are of the order of $ {\cal O}(10^{-8})$, 
whereas they are found to be $ {\cal O}(10^{-7})$
for VLDQ model.

Recently, BaBar Collaboration \cite{babar04} has reported the 
first measurement of the branching ratio for
$B^0 \to \phi \pi^0$ process as $(0.2_{-0.3}^{+0.4} \pm 0.1)\times 10^{-6}$.
Although the error bars are quite large, but this preliminary experimental 
value is almost 2 orders larger than the standard model prediction. 
If in future the data will remain in this order, i.e.,
$ {\cal O}(10^{-7})$, then it will give a clear signal of new physics
effect present in the penguin induced process $B \to \phi \pi$.
As shown, only the VLDQ model can predict such a large branching 
ratio. Therefore, the future experimental data on $B \to \phi \pi$ will
serve as a very good hunting ground for the existence of
new physics beyond the SM   and also support/rule out some of the 
existing new physics models. 
 
\section{Acknowledgement}
We thank Y. Grossman for useful discussions and valuable
comments.
R.M. would like to thank the HEP theory group at the Technion 
for the warm hospitality.
The work of AKG was supported by Lady Davis fellowship and the work of RM was
partly 
supported by Department of Science and Technology, Government of India,
through Grant No. SR/FTP/PS-50/2001.

\end{document}